\documentclass[11pt]{article}
\usepackage{amssymb,amsmath}
\usepackage{slashed}
\usepackage{mathtools}
\usepackage{feynmf}

\usepackage[T1]{fontenc}
\usepackage[utf8]{inputenc}
\usepackage[english]{babel}

\usepackage{mathrsfs}
\usepackage{graphicx}
\usepackage{simplewick}
\usepackage{simpler-wick}
\usepackage{amsxtra}

\usepackage{footnote}
\usepackage{tablefootnote}
\usetikzlibrary{graphs,quotes,arrows.meta}

\usepackage[export]{adjustbox}
\usepackage[colorinlistoftodos]{todonotes}
\usepackage[colorlinks=true, allcolors=blue]{hyperref}
\usepackage{hyperref}
\usepackage{authblk}

\usepackage{tikz}
\usetikzlibrary{graphs}
\usepackage{subfigure}
\usepackage{indentfirst}
\usepackage{textcomp}
\usepackage{gensymb}
\usetikzlibrary{shapes.geometric, arrows}
\usepackage{csquotes}
\usepackage[style=nature,sorting=none,language=autobib,autolang=other,doi=false,url=false]{biblatex}
\graphicspath{{Pics/}}

\title{
Theoretical indication of a possible asymmetry in gamma-radiation between Andromeda halo hemispheres due to Compton scattering on electrons from their hypothetical sources in the halo}
\author{K.M. Belotsky\\ k-belotsky@yandex.ru\\E.S. Shlepkina\\shlepkinaes@gmail.com\\M.L. Soloviev\\max07s@mail.ru}
\affil{National Research Nuclear University MEPhI (Moscow Engineering Physics Institute), 115409, Kashirskoe shosse 31, Moscow, Russia}

\begin{document}
\maketitle

\begin{abstract}

Dark matter (DM) can give observable effects decaying or annihilating with production of electrons or/and photons. Such probability was widely researched for our Galaxy. Here we consider one aspect of similar effect for Andromeda galaxy. We explicitly estimate the energy of the photon of the medium experiencing Inverse Compton (IC) scattering off electron in halo. These photons can be registered by different experiments. Dark matter annihilation or decay could be the source of high energy electrons in halo, though the source could be of other origins too (e.g. running neutron stars). 
Because of specifics in space orientation of Andromeda galaxy disk (a little inclined to the line of sight), the difference in energies could arise for the photons from two hemispheres of Andromeda halo. It is obtained that such asymmetry can be at the level of several 10\%.
\end{abstract}


\section{Introduction}\label{s:intro}

Dark matter (DM) can be the source of high energy electrons and photons due to its annihilation or decay. There are many works elaborating possible observational effects from it in cosmic rays (CR) in our Galaxy. Here we consider issue of possible observational effect from Andromeda galaxy.

If the source of high energy electrons or positrons in halo exist, the process of Inverse Compton (IC) scattering can happen for photons of the medium -- of star light first of all. It is known that the angle distribution of final photon is anisotropic in this scattering process with respect to momentum of incident photon in the initial electron rest frame. This effect should remain in arbitrary reference frame, and will depend on momenta of electron and photon to be scattered. Since star disk of Andromeda galaxy is little inclined with respect to the line of sight, there will be different predominant scattering angle in 'upper' and 'lower' hemispheres of Andromeda halo.  

There should be two effects: in energy and in flux. Here we consider effect in energy only. There should be effect in flux also, which consists in the difference of the values of the photon flux. 

We evaluate a net effect for two fixed points upper and below galaxy disk, what allows doing further predictions of possible effects. 
We will consider the effects of the Andromeda geometry and the line of sight as well as make calculation of the energy spectrum in our future works. It can wash out effect in part, nonetheless it may remain in to some degree, so a geometry modulation of energetic spectrum can be expected. Considered simple case shows how it works. It gives that three energy intervals exist where effect is different: 
at very low final photon energy the ratio $R$ of energies from upper and lower hemispheres is about unity, at higher energy upto $ m^2/\omega\sim 1$ TeV $R\approx 0.6$, where $m$ is the electron mass, $\omega$ is the initial photon energy, and at even higher energy $R\rightarrow 1$. 
Effect should be observed for any photon energy, here we focus on maximum value of final photon energy, though formula obtained is universal. 

Besides effect in the flux, asymmetry in prompt photons radiation from decay/annihilation process (FSR) \cite{123}, comparison with observation sensitivity and background, comparison with other calculation methods \cite{cirelli2009inverse, moskalenko2000anisotropic,fargion1997inverse, fargion1998inverse} are to be taken into account in future.

Andromeda is a rare galaxy which has been recently observed in gamma by Fermi-LAT satellite experiment \cite{abdo2010fermi,ackermann2017observations}. Ground experiments (like HAWC \cite{abeysekara2013hawc}, HESS \cite{hoischen2017grb}, MAGIC \cite{bastieri2006magic}, LHAASO \cite{aharonian2020prospects}, VERITAS \cite{veritas2010veritas}) do not have so high angle resolution (though it depends on energy and they allowed observing several galaxies) and can register only very high energy photons. Effect we are talking about may manifest at any energy including intermediate and low energy ranges. In connection with it, forthcoming satellite gamma-ray telescope project Gamma-400 \cite{egorov2020dark} with especially high angle resolution is of special importance for similar research. There was attempt to connect possible excess in $\gamma$-rays from Andromeda halo with DM \cite{karwin2020dark}. We suggest general feature of asymmetry related with DM or other sources in halo.




\section{IC photon energy}

We assume that there can be sources of high energy electrons or/and positrons in the halo of Andromeda galaxy. Such assumption is based on the attempts to explain positron anomaly \cite{Adriani:2008zr, PhysRevLett.113.121101, 2016ApJ...819...44A} in CR with the help of DM annihilation or decay in our Galaxy. These attempts inevitably involve an effect in gamma-radiation, which was investigated, in particular by our group \cite{belotsky2017fermi, belotsky2019dampe, belotsky2019Indirect, belotsky2020cosmic}. Production of high energy $e^{\pm}$, firstly, is accompanied by FSR, and, secondly, gives energetic photons as a result of IC scattering of these $e^{\pm}$ on photons of the medium. To have IC effect at high energy, 
photons of star light should be taken since they are most energetic from widespread 
radiations within galaxy.

Also one can note that if dark matter annihilation is indeed the origin of the excess of positrons, then we deal with continuously distributed in space high-energy $ e^{\pm} $ sources, so we can take arbitrary point to consider effect.

Let us consider one arbitrary act of electron (or positron, what does not matter in the framework of QED) and photon scattering.

\begin{figure}[ht]
    \centering
    \subfigure[]{
    \includegraphics[width=0.45\textwidth]{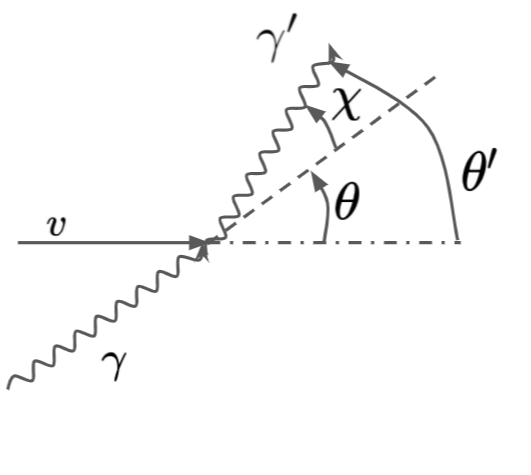}\label{fig:1}}
    \subfigure[]{ \includegraphics[width=0.51\textwidth]{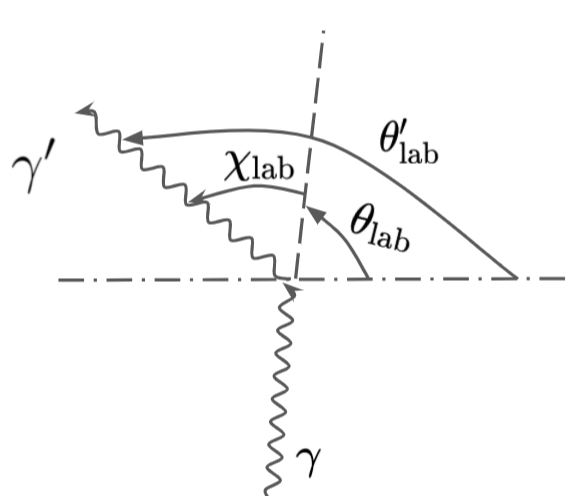}\label{fig:2}}
    \caption{Angle assignment of the scattering process task. Two reference frames are considered: real (a) and of the initial electron rest frame (b). It is seen that in the given angle frame $\theta^{\prime}_{\rm lab}=\theta_{\rm lab}+\chi_{\rm lab}$.}
    \label{fig:1 and 2}
\end{figure}

\begin{figure}[ht]
    \centering
    \includegraphics[width=0.7\textwidth]{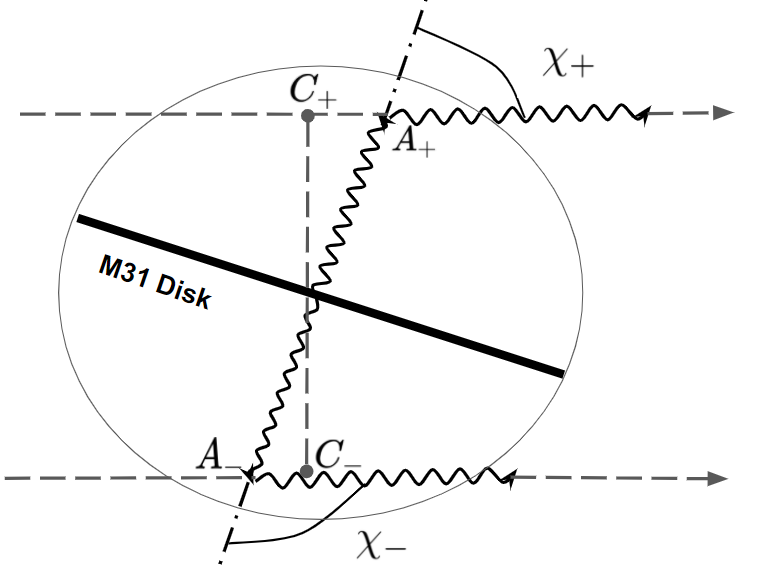}\label{fig:3}
    \caption{The scheme for scattering process in upper and lower Andromeda hemispheres with the chosen points. Observer is on the right.}
\end{figure}

The scheme of the process is shown in the Fig. \ref{fig:1} and \ref{fig:3}. $k$ and $k^{\prime}$ are the initial and final photon 4-momenta, $\omega$ and $\omega^{\prime}$ are their energies, $\theta$ and $\theta^{\prime}$ are the angles between initial momentum of electron and initial and final ones of photons respectively, $\chi$ is the angle between initial and final momenta of the photon. Index 'lab' relates to the same values in the initial electron rest frame.

Let us take Compton formula (one can refer to any textbook, e.g. \cite{peskin2018introduction}) for final photon energy in the 'lab' reference frame
\begin{equation}
   \omega^{\prime}_{\rm{lab}} = \frac{\omega_{\rm{lab}}}{1+ \frac{\omega_{\rm{lab}}}{m}(1 - \cos \chi_{\rm{lab}})},
   \label{eq:1}
\end{equation}
where $m$ is the electron mass, $\chi_{\rm{lab}}$ is the photon scattering angle as shown in the Fig. \ref{fig:2}
We can easily express $\omega_{\rm{lab}}$ from the respective photon energy in the real reference frame ($\omega$) through the Lorentz's transformation:
\begin{equation}
    \omega_{\rm{lab}} = \gamma \, \omega (1 - v\cos \theta),
    \label{eq:2}
\end{equation}
and the same transformation takes place for final photon in the 'lab' frame
\begin{equation}
    \omega_{\rm{lab}}^{\prime} = \gamma \, \omega^{\prime} (1 - v\cos \theta^{\prime}) = \gamma \,\omega^{\prime} (1 - v\cos (\theta + \chi)).
    \label{eq:3}
\end{equation}
We applied relation between angles $\theta^{\prime}=\theta+\chi$ which is seen from the Fig. \ref{fig:1 and 2}. Here and thereafter we use that the absolute value of photon momentum is equal to its energy $\omega^{(\prime)}_{\rm (lab)}$. 
Everywhere $v$ and $\gamma$ mean velocity and $\gamma$-factor of initial electron.

One needs to connect $\cos\chi$ with $\cos \chi_{\rm{lab}}$. It can be done through scalar product of initial and final photon momenta written out in different reference frames and using Lorentz transformations for photon energy:
\begin{equation}
    (kk^{\prime}) = \omega \, \omega^{\prime} (1 - \cos \chi) = \omega_{\rm{lab}} \, \omega_{\rm{lab}}^{\prime}(1 - \cos \chi_{\rm{lab}}).
\end{equation}
From where
\begin{equation}
    1-\cos \chi_{\rm{lab}} = \frac{\omega_{\rm{lab}} \, \omega_{\rm{lab}}^{\prime}}{\omega \, \omega^{\prime}}(1 - \cos\chi),
\end{equation}
where from Eq. \ref{eq:2} and Eq. \ref{eq:3} one gets
\begin{equation}
  \frac{\omega_{\rm{lab}} \,\omega_{\rm{lab}}^{\prime}}{\omega \, \omega^{\prime}} = \frac{1}{\gamma^2 (1-v\cos \theta)(1-v\cos (\theta+\chi))}.
\end{equation}
Substituting in Eq. \ref{eq:1} one obtains
\begin{multline}
 \omega^{\prime}_{\rm{lab}} = \frac{\gamma \, \omega (1 - v\cos \theta)}{1 + \frac{\gamma \, \omega (1 - \cos \theta)}{m}}\frac{1-\cos\chi}{\gamma^2(1 - v \cos \theta)(1 - v \cos (\theta + \chi))} =  \\
  = \frac{\gamma \, \omega (1 - v\cos \theta)}{1+ \frac{\omega}{\gamma \, m}\frac{1 - \cos \chi}{1 - v \cos (\theta + \chi)}}.
  \label{eq:7}
\end{multline}
From Eq. \ref{eq:3} one has
\begin{equation}
    \omega^{\prime} = \frac{\omega^{\prime}_{\rm{lab}}}{\gamma(1 - v\cos (\theta + \chi))},
\end{equation}
and, finally, taking into account Eq. \ref{eq:7} one gets
\begin{equation}
    \omega^{\prime} = \frac{\omega (1 - v \cos \theta)}{1 - v \cos (\theta + \chi) + \frac{\omega}{\gamma \, m}(1 - \cos \chi)},
    \label{final form}
\end{equation}
%
where $\frac{\omega}{m} \equiv\gamma_{cr}^{-1}\sim(2\div 4) \cdot 10^{-6}$. 

One can see from Eq. \ref{final form} that there are different situations that 
can be easily analyzed. Let the velocity be $v\approx 0$.
 Then the small third term in denominator $\omega/m(1-\cos\chi)$ will give a tiny anisotropy between hemispheres (what is not important for us). The energy ratio between upper and lower hemispheres 
 \begin{equation}
    R\equiv \omega^{\prime}_+/\omega^{\prime}_-
\end{equation}
will be a little bigger than unity for the chosen two points $A_+$ and $A_-$ in Fig. \ref{fig:3}, where $\omega^{\prime}_{\pm}$ are the final photon energies from upper and lower hemispheres respectively.
Scattering angles in upper and lower hemispheres of Andromeda $\chi_{\pm}$ are introduced 
in the Fig. \ref{fig:3}.
When $v\sim 1$, third term in denominator of Eq. \ref{final form} (which is proportional to $\omega/m$) is negligible and $\omega^{\prime}$  comes to maximum at $\cos(\theta + \chi) = 1$. So $\theta = -\chi $, what corresponds to the case when initial electron goes in direction to the observer\footnote{One notes that consideration of other configuration of momenta (angles) in the Fig.\ref{fig:1 and 2} \ref{fig:1 and 2} can lead to the relation $\theta^{\prime}=\theta-\chi$. But it does not change conclusion that $\cos(\theta^{\prime})=\cos(\theta-\chi)=1$ and that in the given angle system frame initial electron travels in direction to the observer.}. It corresponds to narrow sharp maximum in photon energy which is of bigger interest. Next, if $v\rightarrow 1$ so $1-v$ becomes smaller than $\omega/\gamma m$, i.e. when $\gamma\gg \gamma_{cr}$, the third term in denominator starts to dominate again and in this limit denominator and numerator are canceled.

Finally, we obtain for maximal final photon energy in the real reference frame
\begin{gather}
    \omega^{\prime}_{\max} = \frac{\omega (1 - v \cos \chi)}{1-v + \frac{\omega}{\gamma \, m}(1 - \cos \chi)} = \frac{(1+v)\gamma^2\omega (1 - v \cos \chi)}{1 + (1+v)\gamma \frac{\omega}{m}( - \cos \chi)} \sim \nonumber \\ \sim
    \begin{cases}
    \frac{1}{1+\frac{\omega}{m}(1 - \cos \chi)} \approx 1 & v = 0\\
    1 - \cos \chi & 1 \ll \gamma \ll \gamma_{cr}\\
    \frac{1 - \cos\chi}{1 - \cos \chi} = 1 & \gamma \gg \gamma_{cr}.
    \end{cases}
    \label{eq:11}
\end{gather}
So, there exists wide electron energy interval, $1 \ll \gamma \ll \gamma_{cr}$ (what corresponds to initial electron energy $m\ll E\ll $ TeV for $\omega\sim 1$ eV), where the effect takes place 
\begin{equation}
    R=\frac{1-\cos\chi_+}{1-\cos\chi_-}\approx 0.6
\end{equation}
for the chosen two points $A_+$ and $A_-$ in the Fig. \ref{fig:3}.

\begin{figure}[!ht]
    \centering
    \includegraphics[width=0.7\textwidth]{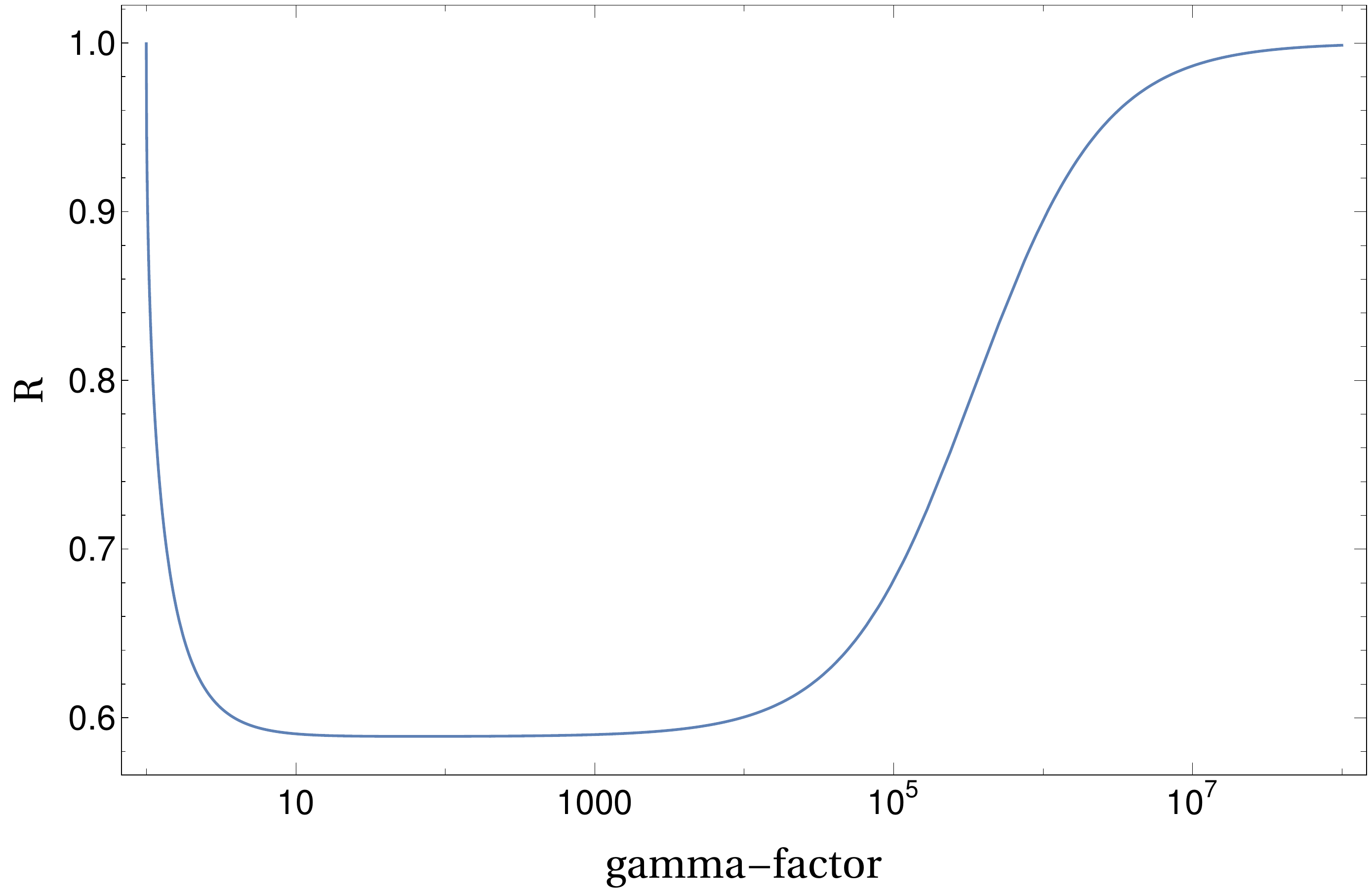}
    \caption{Dependence of $R$ showing asymmetry effect for maximal observed photon energy $\omega^{\prime}_{\text{max}}$ between upper and lower hemispheres of Andromeda galaxy from $\gamma$-factor of initial electron. Figure relates to the chosen two points $A_+$ and $A_-$ of Fig. \ref{fig:3}}
    \label{fig:4}
\end{figure}

The ratio $R$ for maximal photon energy as dependent on $\gamma$-factor followed from the Eq. \ref{eq:11} is illustrated in Fig. \ref{fig:4}.

\vspace{0.5 cm}
\textit{Comments on line of sight integration }
\vspace{0.5 cm}

We have shown explicit effect for two fixed points above and below galaxy disk. Certainly, one should take into account the effects caused by our line of sight and distributed photon and electron sources, what could wash out difference between hemispheres. Nonetheless effect should remain in some degree since situation for upper line of sight and lower one is not symmetric (see Fig.\ref{fig:3}) because of dependence of Compton scattering from initial relative angle between momenta of the scattered electron and photon. As was seen from Eq. \ref{eq:11}, when initial photon and electron move towards each other ($\theta>90^{\circ}$ in Fig. \ref{fig:3}) final photon (maximal) energy $\omega^{\prime}$ is bigger in wide its value range than when they move co-directionally  ($\theta<90^{\circ}$).

Line of sight Integration can be done taking into account a flux and under extra assumption about $e^{\pm}$ source distribution. Qualitatively, two factors will make difference between upper and lower lines of sight: it is density of $e^{\pm}$ sources and density of medium photons. Density of the sources is expected to decrease from distance to the galaxy center, concentration of photons - from distance to the stars in disc. For example, one can consider two nearest to the galactic center points $C_+$ and $C_-$ in Fig. \ref{fig:3} of two opposite lines of sight. Source densities in them is expected to be equal and maximal over all given lines of sight. But in point $C_+$ the closest part of galaxy disc will shine co-directionally, while for $C_-$ will do towards. Similarly one can consider other parts of lines of sight and there will be effects of different signs, though their full compensation is hardly expected.

\section{Conclusion}
We considered possible effect of asymmetry of gamma radiation from another galaxy connected with its geometric orientation with respect to line of sight.Here we considered an effect in energy between two hemispheres of galaxy related to IC scattering of medium photons on high energy $e^{\pm}$ from their hypothetical source in galaxy halo. There also will be an effect in flux. The effect can be achievable  for existing or future experiments since it may not seem to be vanishing, it can be at the level 10 (several tens) percents.

Our future work will concern the flux, sensitivity and data on cosmic gamma-background, and include the prompt (FSR) photons appearing under the assumption  that $e^{\pm}$ has DM decay/annihilation origin.

\section*{Acknowledgements}
The work was supported by the Ministry of Science and Higher Education of the Russian Federation
by project No 0723-2020-0040 ``Fundamental problems of cosmic rays and dark matter''. Also we would like to thank A.Egorov, M.Khlopov, M.Laletin and S.Rubin for interest to the work and useful discussions and also D.Fargion for providing relevant references.


\printbibliography

\end{document}